\documentclass[11pt]{article}
\usepackage{vietnam,epsfig}

\newcommand{\dzero}     {D\O\  }
\newcommand{\ppbar}     {\mbox{$p\bar{p}$}\ }
\newcommand{\met}       {\mbox{$\not\!\!E_T$}}
\newcommand{\ttbar}      {\mbox{$t\bar{t}$}\ }

\def\be{\begin{equation}}
\def\ee{\end{equation}}
\def\bea{\begin{eqnarray}}
\def\eea{\end{eqnarray}}

\begin{document}
\vspace*{4cm}
\title{Top Quark Production Cross Section at $E^{cm}=1.96$TeV}

\author{ Reinhard Schwienhorst\footnote{On behalf of the \dzero and CDF collaborations.} }

\address{Department of Physics and Astronomy, Michigan State University, East Lansing, MI
48824. \\
        Email: schwier@fnal.gov}

\maketitle\abstracts{
Preliminary results on the top pair production cross section measurements by the
\dzero and CDF experiments in Run~II at the Tevatron are presented. The measurements
are obtained using various final state signatures.}

\section{Introduction}
The top quark was discovered by the \dzero and CDF collaborations in 
1995\cite{Abe:1995hr,Abachi:1995iq} at the Tevatron \ppbar collider. Top quarks are 
produced mainly in pairs through the strong interaction at the Tevatron. The Standard Model
cross section including NNLO soft-gluon corrections is 
$\sigma_{\ttbar}=6.77\pm 0.47$pb\cite{Kidonakis:2003qe}.
The decay of the two top quarks, $t \rightarrow W b$, leads to three distinct final state signatures, 
depending on the decay products of the $W$. The lepton+jets
channel where one of the $W$'s decays to an electron or a muon and the other one decays 
hadronically contains about 30\% of the \ttbar events. The di-lepton channel where both $W$'s decay
to an electron or a muon contains only about 5\% of the \ttbar events, but it also has small 
backgrounds. The all-hadronic channel where both $W$'s decay hadronically contains 44\% of the 
\ttbar events but it also suffers from very large backgrounds due to QCD multi-jet production. 
Tau decays of the $W$ are not considered here.

In this paper we present preliminary results for the cross section measurement 
from the \dzero and CDF experiments in the lepton+jets, di-lepton, and all-hadronic channels
in Run~II at the Tevatron with a center-of-mass energy of 1.96TeV.


\section{Di-lepton Channel}
Top pair decays to di-leptons provide a clean signature with small backgrounds from di-boson production 
and $Z\rightarrow \tau\bar{\tau}$ events as well as $Z/\gamma$+jets events with mis-measured \met.
The main contributions to the systematic uncertainty are from jet energy scale, object ID, as well as background
normalization. Modeling of b-tagging in the Monte Carlo also contributes to the systematic uncertainty.

\begin{figure}
\begin{minipage}{0.5\textwidth}
  \centering
  \psfig{figure=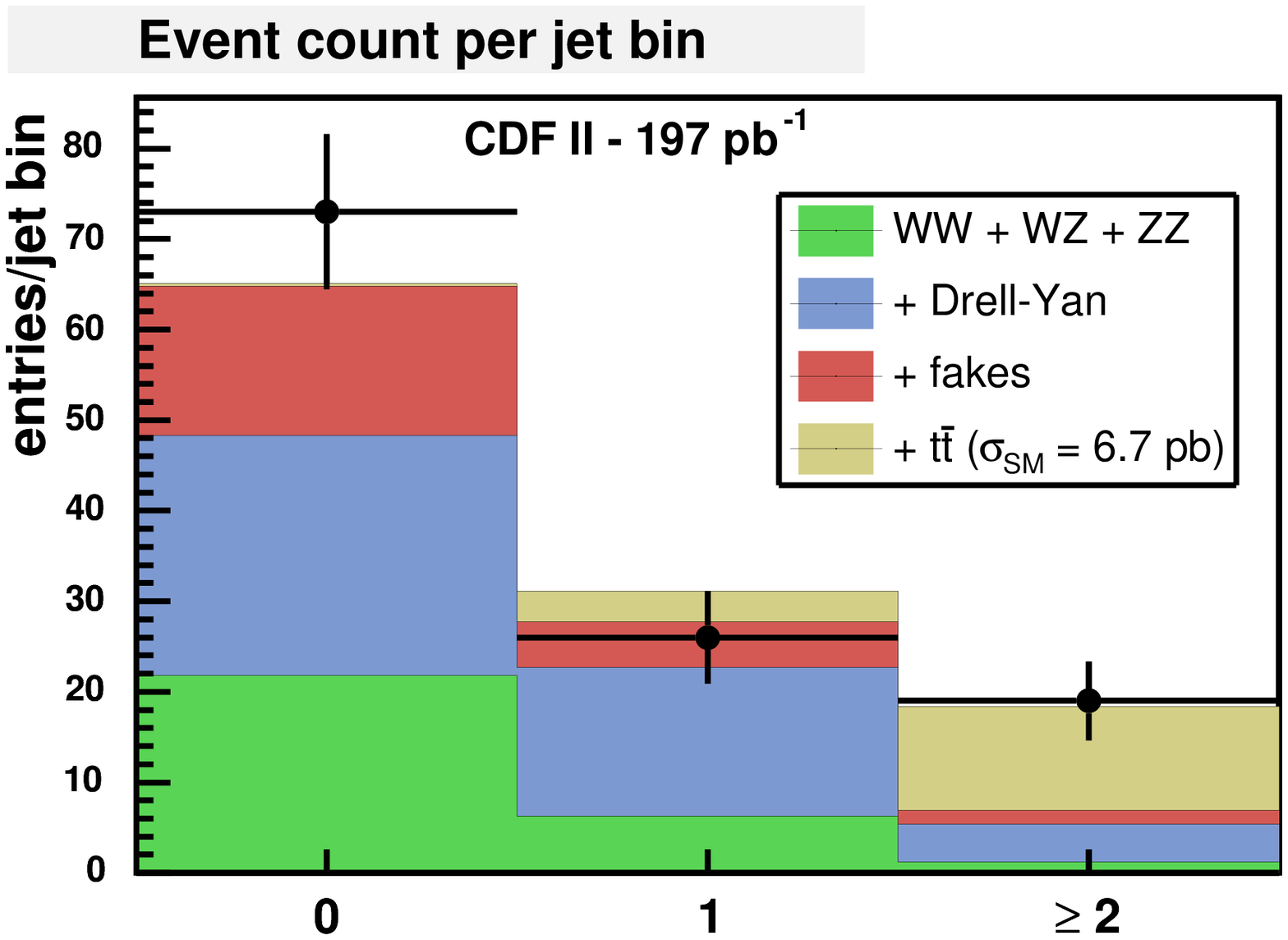,height=2.2in}
\end{minipage}
\begin{minipage}{0.5\textwidth}
  \centering
  \psfig{figure=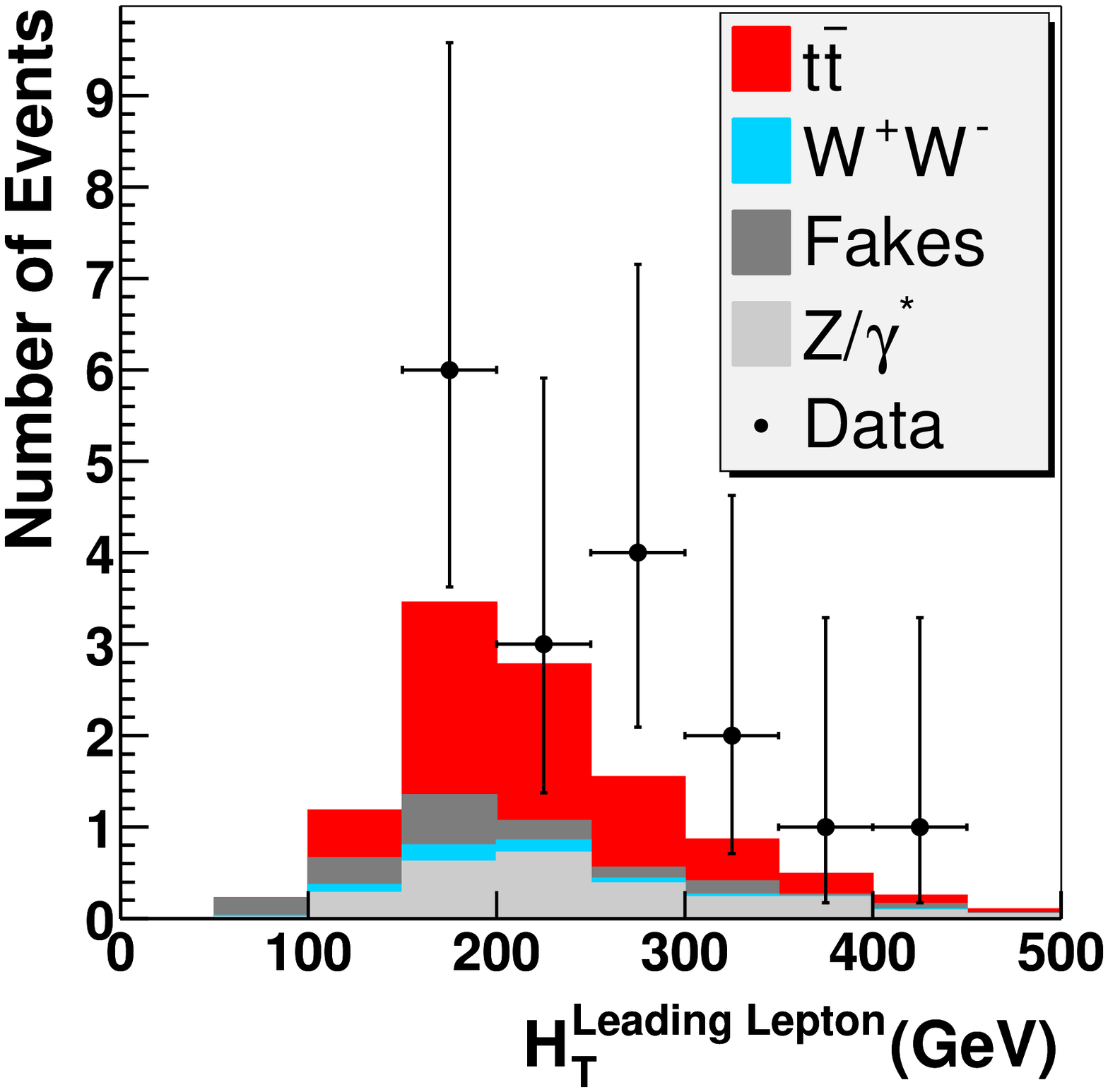,height=2.5in}
\end{minipage}
\caption{The jet multiplicity distribution for the CDF lepton+track analysis (left) 
and the $H_T^{\rm leading\; lepton}$ distribution for the \dzero di-lepton analysis (right).
\label{fig:dilep}}
\end{figure}

The CDF experiment uses several different approaches to measure the \ttbar cross section in the di-lepton
channel in 200pb$^{-1}$ of Run~II data. The lepton+track analysis asks for one electron or muon to be well 
identified in the detector with $p_T>20GeV$ and $|\eta|<1$ (muons) or $|\eta|<2$ (electrons).  
The requirement on the additional lepton is only to ask for one isolated track with $p_T>20GeV$, 
thus maximizing the acceptance for di-lepton events and even including some $W\rightarrow\tau$ events. Further
requirements are $\met>25GeV$ and at least two jets with $E_T>20GeV$ and $|\eta|<2$. The resulting yield
is compared to the sum of the background contributions plus the expected Standard Model
\ttbar signal in the plot on the left-hand side of Fig.~\ref{fig:dilep}. The measured cross section,
including systematic uncertainties, is 
$\sigma_{\ttbar} = 7.0^{+2.7}_{-2.3}(stat)^{+1.5}_{-1.3}(syst)\pm 0.4(lumi)$pb.
A cut-based analysis has also been performed, requiring two oppositely charged, well-identified leptons
with $p_T>20GeV$, $\met>20GeV$, at least two jets, and total transverse energy of the event $H_T>200GeV$, 
where $H_T$ is the scalar sum of the transverse energies of the lepton, \met, and the jets.
The measured cross section for this analysis is 
$\sigma_{\ttbar} = 8.4^{+3.2}_{-2.7}(stat)^{+1.5}_{-1.1}(syst)\pm 0.5(lumi)$pb.
In a third analysis, a likelihood fit has been performed in the $\met-n_{jets}$ plane to determine 
simultaneously  the contributions from \ttbar, $WW$, and $Z\rightarrow \tau\bar{\tau}$. 
The measured cross section
for \ttbar is $\sigma_{\ttbar} = 8.6^{+2.5}_{-2.4}(stat)\pm 1.1(syst)$pb.

The \dzero experiment has done separate analyses in the $ee$, $e\mu$, and $\mu\mu$ final states in about 
140pb$^{-1}$ of Run~II data. The analyses are requiring two isolated leptons (electron
or muon) with $p_T>15GeV$ ($p_T>20GeV$ for the $ee$ channel) and $|\eta_{det}|<1.1$ for electrons and 
$|\eta_{det}|<2.0$ for muons. Additional requirements are $\met>25GeV$ for the $e\mu$ channel and $\met>35GeV$
for the $ee$ and $\mu\mu$ channel as well as at least two jets with $E_T>20GeV$. Events in which the invariant
mass of the $ee$ or $\mu\mu$ pair is consistent with a $Z$ boson are removed. To separate signal events from
the backgrounds, a cut is made on the transverse event energy $H_T^{\rm leading\; lepton}>120GeV$ for the 
$\mu\mu$ channel and $H_T^{\rm leading\; lepton}>140GeV$ for the $e\mu$ channel, where 
$H_T^{\rm leading\; lepton}$ is the scalar sum of the transverse energies of the jets, the leading lepton, 
and \met. The right-hand side of Fig.~\ref{fig:dilep} shows a comparison between the sum of the background 
contributions and the expected Standard Model \ttbar signal to the data for $H_T^{\rm leading\; lepton}$.
The measured cross section is 
$\sigma_{\ttbar} = 14.3^{+5.1}_{-4.3}(stat)^{+2.6}_{-1.9}(syst)\pm 0.9(lumi)$pb.
The \dzero experiment has also performed a measurement in the $e\mu$ channel requiring that one of the jets is
b-tagged with a secondary-vertex tagger. As a result this channel is virtually background-free but has low 
statistics. The measured cross section using 140pb$^{-1}$ of Run~II data is 
$\sigma_{\ttbar} = 11.1^{+5.8}_{-4.3}(stat) \pm 1.1(syst)\pm 0.6(lumi)$pb.


\section{Lepton+Jets Channel}
The \ttbar decay mode where one of the $W$'s decays to an electron or muon and the other $W$ to quarks results 
in a final state of one lepton and at least three high-$E_T$ jets. The dominant background to this event 
signature is from $W/Z$+jets production, and an additional background comes from QCD multi-jet events where 
one of the jets is mis-identified as an isolated lepton.

\begin{figure}
\begin{minipage}{0.5\textwidth}
  \centering
  \psfig{figure=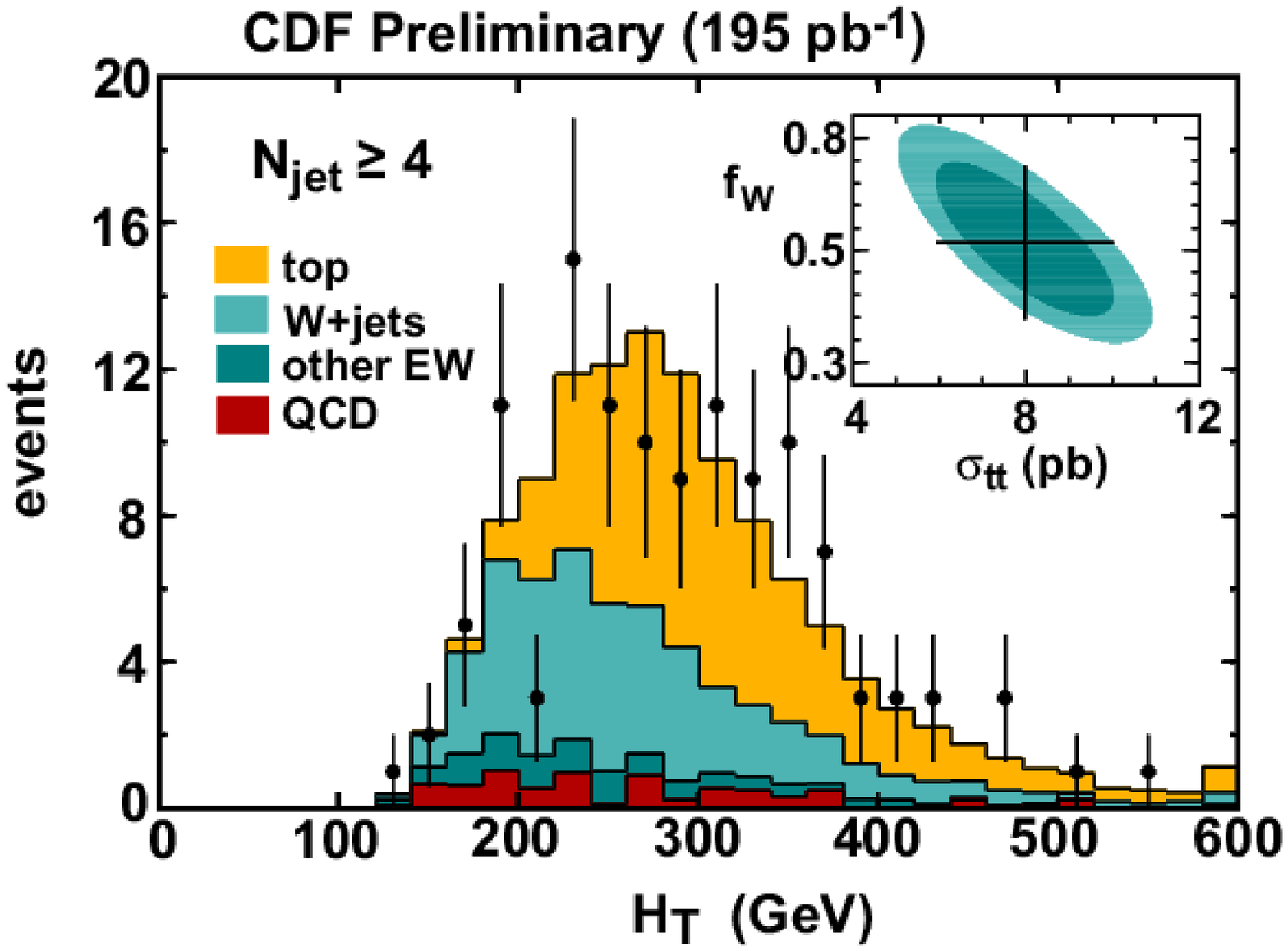,height=2.2in}
\end{minipage}
\begin{minipage}{0.5\textwidth}
  \centering
  \psfig{figure=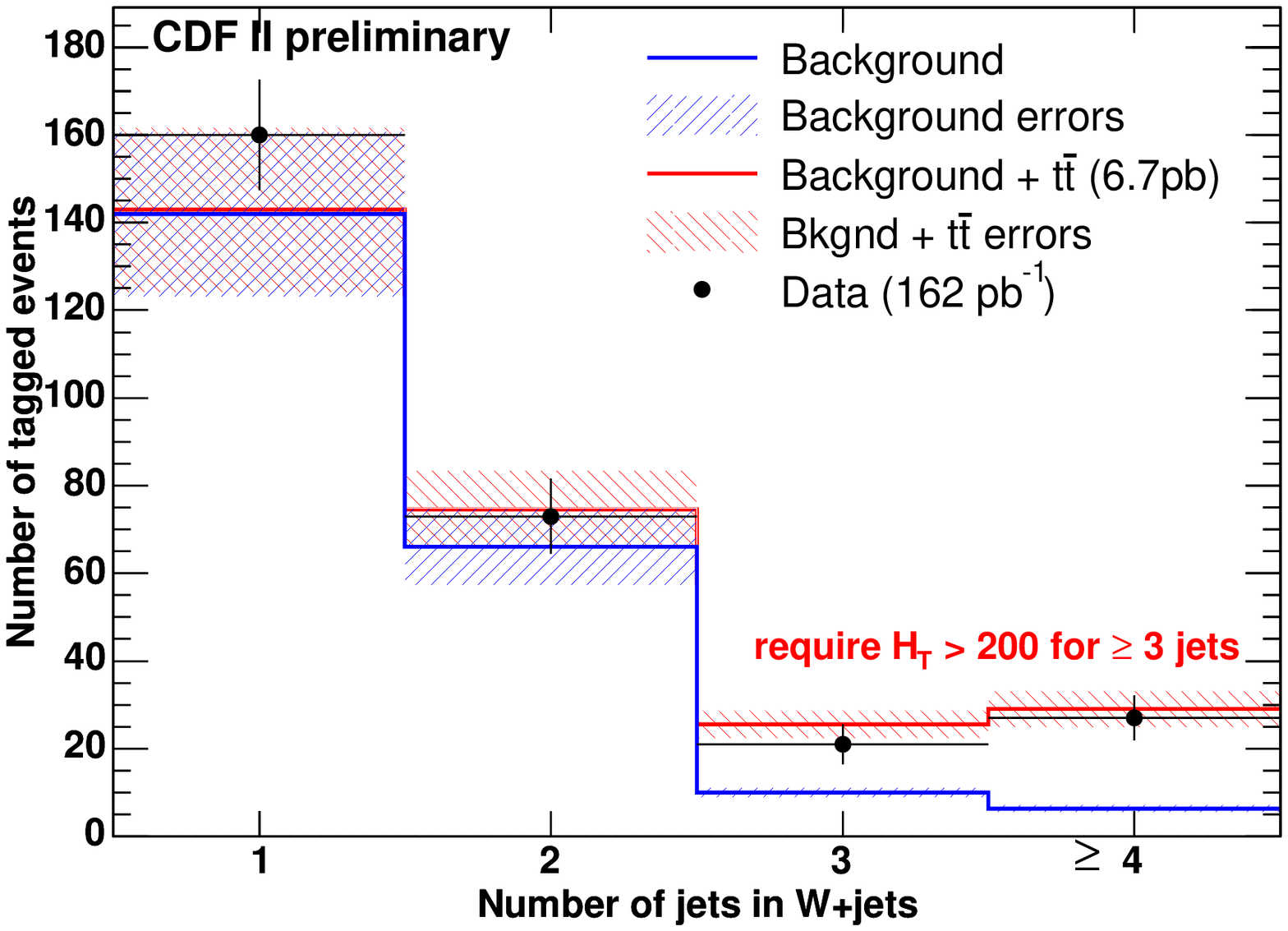,height=2.2in}
\end{minipage}
\caption{The $H_T$ distribution for the CDF lepton+jets analysis (left) and the jet multiplicity
distribution for the CDF lepton+jets analysis with b-tagging (right).
\label{fig:cdflepjets}}
\end{figure}

The CDF experiment has done several separate analyses using 200pb$^{-1}$ of Run~II data, all using the same 
basic selection cuts: exactly
one isolated lepton (electron or muon) with $p_T>20GeV$, $\met>20GeV$, and at least three jets with
$E_T>15GeV$. One analysis requires at least four jets and performs a template fit of the background and 
signal $H_T$ distributions to the observed data. The result is shown on the left-hand side of 
Fig.~\ref{fig:cdflepjets}. The measured cross section in this analysis is 
$\sigma_{\ttbar} = 4.7\pm 1.6(stat)\pm 1.8(syst)$pb.
Another analysis uses a 7-input Neural Network to separate signal from background and performs a template
fit to the Neural Network output, giving a measured cross section of 
$\sigma_{\ttbar} = 6.7\pm 1.1(stat)\pm 1.5(syst)$pb.

CDF has also performed two analyses in the lepton+jets channel with 200pb$^{-1}$ of Run~II data, 
requiring that at least one of the jets is
tagged with a secondary vertex b-tagging algorithm. This additional requirement reduces the backgrounds
significantly. One of the b-tagging analyses makes the additional requirement that $H_T>200GeV$. The 
resulting jet multiplicity distribution is shown on the right-hand side of Fig.~\ref{fig:cdflepjets}.
The measured cross section for events with at least one b-tag is
$\sigma_{\ttbar} = 5.6^{+1.2}_{-1.1}(stat)^{+0.9}_{-0.6}(syst)$pb.
The measured cross section for events with at least two b-tag and no cut on $H_T$ is
$\sigma_{\ttbar} = 5.0^{+2.4}_{-1.9}(stat)^{+1.1}_{-0.8}(syst)$pb.
The other b-tagging analysis performs a template fit to the $E_T$ distribution of the leading jet in
the event to separate signal from backgrounds, resulting in a measured cross section of
$\sigma_{\ttbar} = 6.0^{+1.5}_{-1.8}(stat)\pm 0.8(syst)$pb.

\begin{figure}
\begin{minipage}{0.5\textwidth}
  \centering
  \psfig{figure=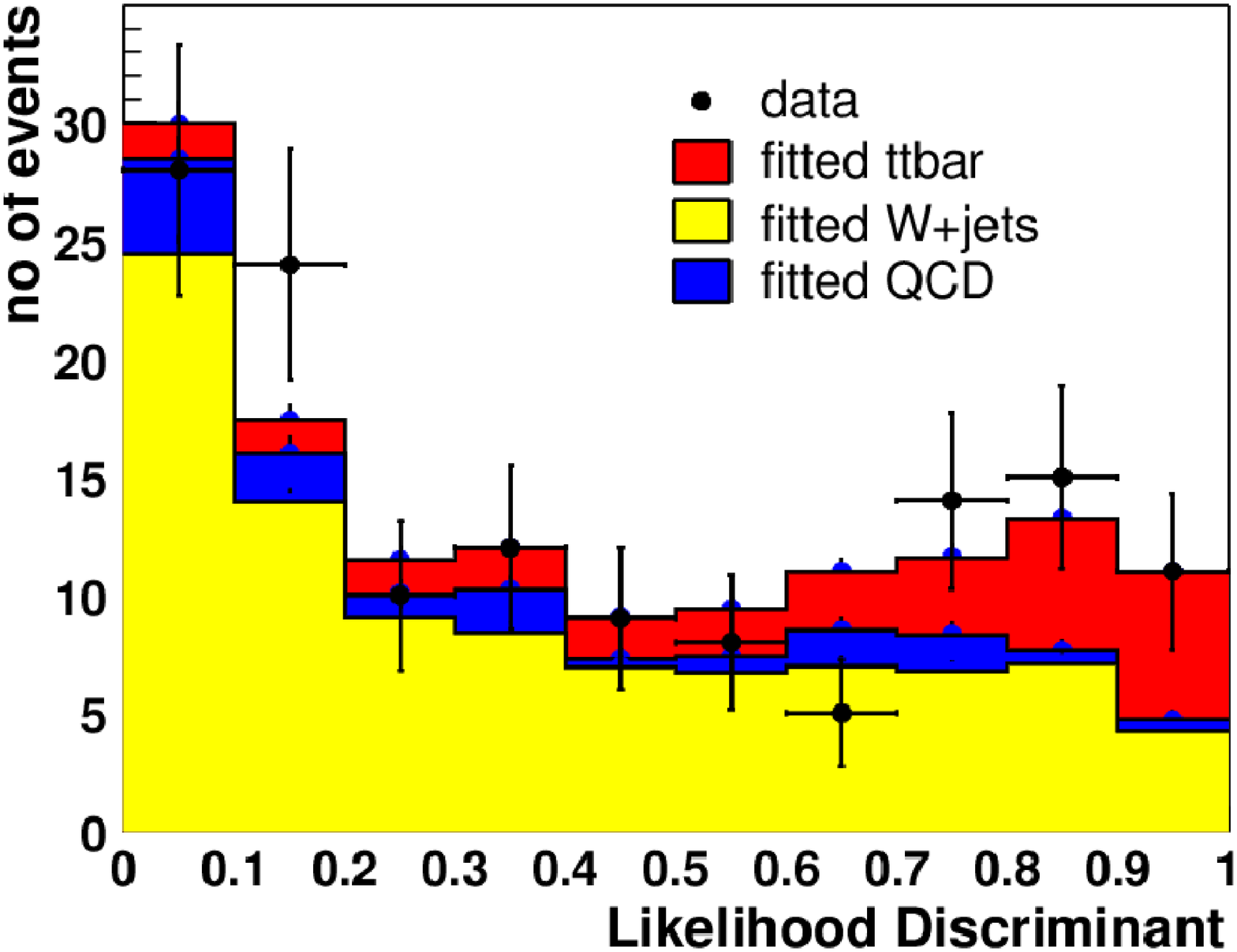,height=2.2in}
\end{minipage}
\begin{minipage}{0.5\textwidth}
  \centering
  \psfig{figure=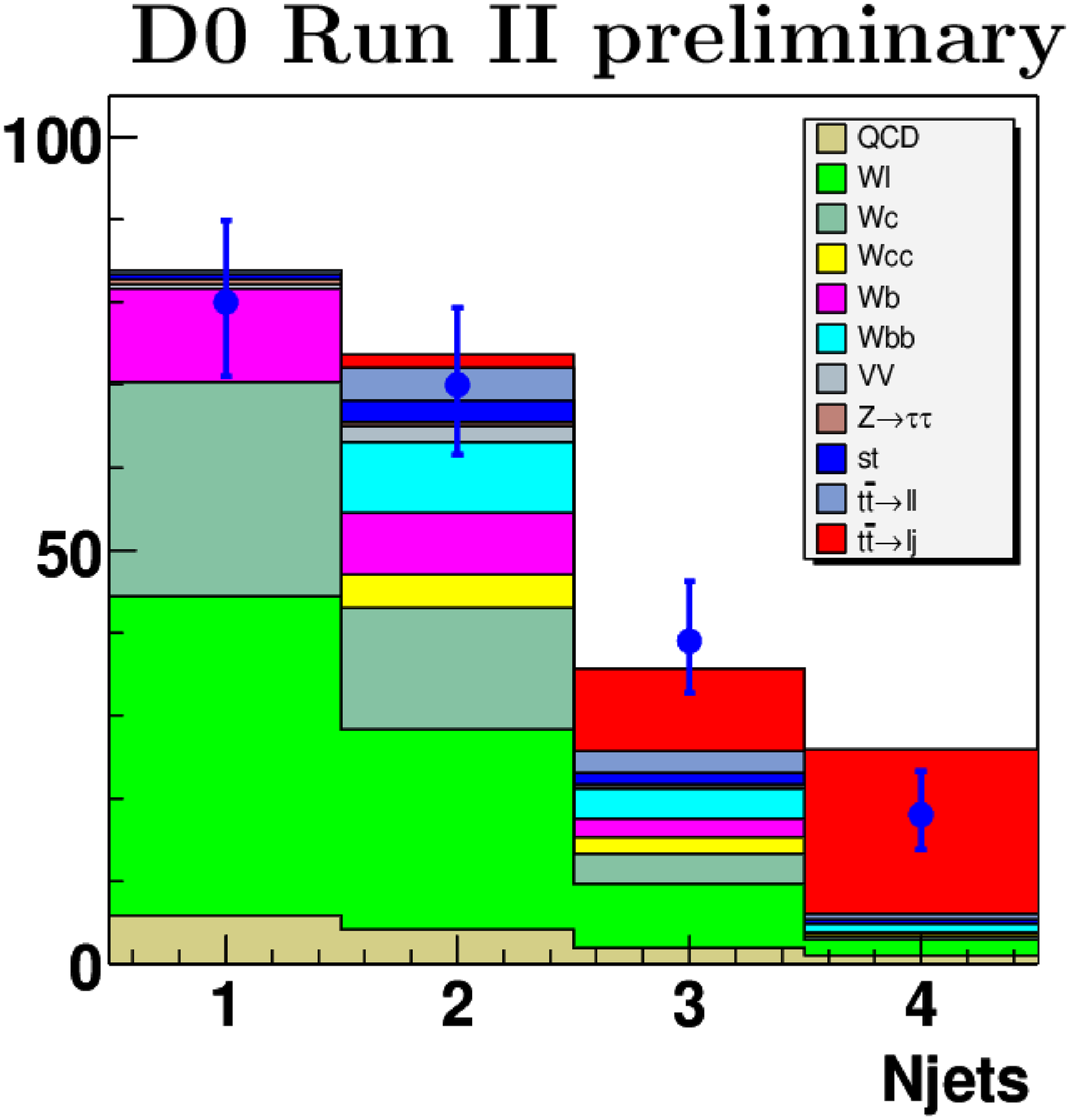,height=2.2in}
\end{minipage}
\caption{The result of the template fit to the likelihood distribution in the \dzero electron channel 
analysis (left) and jet multiplicity for the muon channel with CSIP tag analysis (right).
\label{fig:d0lepjets}}
\end{figure}

The \dzero experiment has performed several analyses in the lepton+jets channel using 160pb$^{-1}$
of Run~II data. The same basic selection cuts are used for all lepton+jets analyses: 
one isolated lepton (electron or muon) with $E_T>20GeV$, $\met>20GeV$ ($\met>17GeV$)
for the electron (muon) channel, and at least three jets with $E_T>15GeV$ and $|\eta|<2.5$.
A topological analysis requires at least four jets and then builds a likelihood discriminant from 
four topological variables that each give good signal-background separation. A template fit to the 
likelihood discriminant is then performed to obtain a cross section measurement. The likelihood discriminant
for the electron channel data is shown in the left-hand plot of Fig.~\ref{fig:d0lepjets} 
together with the fit result. The measured cross section is
$\sigma_{\ttbar} = 7.2^{+2.7}_{-2.4}(stat)^{+1.6}_{-1.7}(syst)\pm 0.5(lumi)$pb.

The \dzero experiment has also performed an analysis in the muon channel using two different
tagging algorithm, a) based on the reconstruction of a secondary vertex (SVT), and b) based on
measuring the impact parameter of tracks with respect to the primary interaction vertex (CSIP). 
The jet multiplicity distribution
for CSIP tagged events in data is shown on the right-hand side of Fig.~\ref{fig:d0lepjets},
together with the individual background and signal contributions. The measured cross section for the
CSIP tagger is $\sigma_{\ttbar} = 5.2^{+1.7}_{-1.5}(stat)^{+1.7}_{-1.2}(syst)\pm 0.3(lumi)$pb. The
measured cross section using the SVT tagger is
$\sigma_{\ttbar} = 6.9^{+2.0}_{-1.8}(stat)^{+1.8}_{-1.7}(syst)\pm 0.4(lumi)$pb.


\section{All-Hadronic Channel}
The all-hadronic decay mode is a more difficult measurement because it has a large background from 
QCD multi-jet production. This background is suppressed by requiring at least one of the jets to be
b-tagged.
The CDF experiment selects events with at least six jets, requires at least one b-tag, and makes 
several cuts on kinematical variables to reduce the multi-jet background further. The measured
cross section is $\sigma_{\ttbar} = 7.8\pm 2.5(stat)^{+4.7}_{-2.3}(syst)$pb.
The \dzero experiment selects events with at least six jets and requires at least one secondary
vertex b-tag. A neural network is then constructed using several kinematical variables. The 
cross section is measured in a fit to the neural network output. The measured cross section is
$\sigma_{\ttbar}=7.7^{+3.4}_{-3.3}(stat)^{+4.7}_{-3.7}(syst)\pm 0.5(lumi)$pb.


\section{Summary}
The top pair production cross section has been measured in Run~II at the Tevatron by both the
\dzero and CDF collaborations in various final states using several different approaches, 
in datasets of 140pb$^{-1}$ to 200pb$^{-1}$.
The statistical uncertainty is comparable to the systematic uncertainty in many analysis channels. 
The measured cross sections are consistent with the Standard Model expectation.


\end{document}